\documentclass[twocolumn,showpacs,prl]{revtex4}
\usepackage{amssymb}
\usepackage{amsmath}
\usepackage{graphicx}
\usepackage{dcolumn}
\usepackage{bm}

\begin{document}

\title{Tensor network simulation of phase diagram of frustrated $J_{1}$-$%
J_{2}$ Heisenberg model \\
on a checkerboard lattice }
\author{Y.-H. Chan$^{1,2}$, Y.-J. Han$^{1,3}$, and L.-M. Duan$^{1,2}$}
\affiliation{$^{1}$Department of Physics and MCTP, University of Michigan, Ann Arbor,
Michigan 48109, USA}
\affiliation{$^{2}$Center for Quantum Information, IIIS, Tsinghua University, Beijing,
China}
\affiliation{$^{3}$Key laboratory of Quantum Information, University of Science and Technology of China, Hefei, Anhui 230026, People's Republic of China}

\begin{abstract}
We use the recently developed tensor network algorithm based on infinite
projected entangled pair states (iPEPS) to study the phase diagram of
frustrated antiferromagnetic $J_{1}$-$J_{2}$ Heisenberg model on a
checkerboard lattice. The simulation indicates a Neel ordered phase when $%
J_{2}<0.88J_{1}$, a plaquette valence bond solid state when $%
0.88<J_{2}/J_{1}<1.11$, and a stripe phase when $J_{2}>1.11J_{1}$, with two
first-order transitions across the phase boundaries. The calculation shows
the cross-dimer state proposed before is unlikely to be the ground state of
the model, although such a state indeed arises as a metastable state in some
parameter region.
\end{abstract}

\maketitle

Understanding frustrated quantum magnetic models is a long standing
difficult problem in strongly correlated physics. Theoretical tools to study
these systems are limited. Exact diagonalization is limited by the small
system size, and quantum Monte-Carlo simulation is hindered by the infamous
sign problem. Among the frustrated models, the antiferromagnetic $J_{1}$-$%
J_{2}$ Heisenberg model on a checkerboard lattice (or called the crossed
chain model) is an important example that has raised a lot of interest \cite%
{1,1',2,3,4,5,6}, due to its rich phase diagram and connection with real
materials. This model is described by the Hamiltonian
\begin{equation}
H=J_{1}\sum_{\left\langle i,j\right\rangle }\vec{S_{i}}\cdot \vec{S_{j}}%
+J_{2}\sum_{\left\langle \langle i,j\right\rangle \rangle }\vec{S_{i}}\cdot
\vec{S_{j}}
\end{equation}%
where $J_{1}$ is the nearest-neighbor spin coupling rate on a square lattice
and $J_{2}$ is the next-nearest-neighbor spin coupling rate on a
checkerboard pattern of plaquettes, as depicted in Fig. 1. A number of works
provide complimentary studies in different parameter regions. The complete
phase diagram for this model, however, still remains controversial. It is
known that the system has a collinear long-range Neel order when $%
J_{2}<<J_{1}$. At $J_{1}\approx J_{2}$, the calculation based on the
strong-coupling expansion predicts a plaquette valence bond solid as the
ground state \cite{1,2,3,4}. In the region with $J_{2}>J_{1}>0$, the phase
is still under debate. Possibilities include the fourfold degenerate state
with long range spin order, supported with semi-classical studies \cite{5}
and large-$N$ expansion calculation \cite{6}, the sliding Luttinger liquid
phase, supported with perturbative random phase calculation and exact
diagonalization of a small system \cite{1}, and the cross dimer state,
supported with bosonization approach \cite{2} and two-step DMRG
(density-matrix renormalization group) simulation \cite{3}.

Recently, tensor network algorithms emerge as a promising method to solve
two-dimensional frustrated quantum systems \cite{7,10,11,12,13}. There are
different types of tensor network algorithms, but all the algorithms share
the basic idea to describe the ground state of the model Hamiltonian as a
tensor network state that respects the entanglement area law. The tensor
network algorithms belong to the variational method and have no intrinsic
sign problem for frustrated systems. The tensor network algorithms have been
tested for a number of non-frustrated Hamiltonians, and the results agree
pretty well with quantum Monte Carlo simulation \cite{11}. Recently, the
algorithms have also been applied to the frustrated Heisenberg model on a
Kagome lattice \cite{12} and the $J_{1}$-$J_{2}$-$J_{3}$ model on a square
lattice \cite{13}.

In this paper, we use a particular type of tensor network algorithm, the
iPEPS (infinite PEPS) \cite{10}, to simulate frustrated antiferromagnetic $%
J_{1}$-$J_{2}$ model on a a checkerboard lattice in the thermodynamic limit.
We construct the complete phase diagram with the following findings: (1) the
simulation shows two first-order phase transitions respectively at $%
J_{2}/J_{1}=0.88$ and $J_{2}/J_{1}=1.11$, first from a Neel state to a
plaquette valence bond solid, and then to a spin ordered stripe phase. (2)
In the region with $J_{2}/J_{1}>1.11$ (except for the special point $J_{1}=0$%
), our calculation supports the four-fold degenerate states proposed in Ref.
\cite{5} as the ground state. In particular, the spin stripe phase seems to
be the most stable one under perturbation. (3) Our simulation provides
strong evidence to show that the cross-dimer state is not the ground state
of the system, although it indeed emerges as a metastable state in some
parameter region (its energy is always significantly higher compared with
the spin stripe phase).

\begin{figure}[tbp]
\includegraphics[width=6cm,height=3cm]{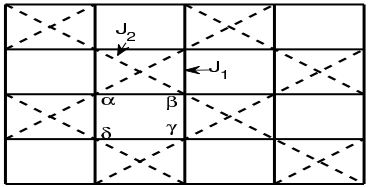}
\caption[Fig. 1 ]{Illustration of the $J_{1}-J_2$ Heisenberg model on a checkerboard lattice.}
\end{figure}

In the iPEPS algorithm, at every site, we represent the state as a
five-index tensor with one physical index (with dimension two for a
spin-half system) and four virtual indices (with internal dimension denoted
by $D$) \cite{10}. The wave function can be obtained by contracting all the
virtual indices. To obtain the expectation value of a physical quantity, we
need to first contract the physical index to form a tensor network with
internal dimension $D^{2}$, and this tensor network is then contracted from
the infinite boundary through multiplication with a matrix product state
with internal dimension $\chi $. We tune the value of $\chi $ until
convergence is achieved in the calculated physical quantity. As a thumb of
rule, typically at $\chi \gtrsim D^{2}$,\ the relative error of energy
induced by variation of $\chi $ has been reduced to the order of $10^{-5}$,
which indicates good convergence already. The dominant error of the
calculation is from the small value of the internal dimension $D$.
As the calculation time scales with $D$ as $D^{12}$ (under choice of $\chi
\sim D^{2}$) \cite{10}, the value of $D$ in our simulation is limited to be about $5$.
We compare the energy as well as several other quantities (including the
phase boundary specified below in Fig. 2) calculated with $D=4$ and $D=5$,
and the difference is within a percent level. As an estimate, we expect that
the relative error of our numerical simulation, dominated by the limited
value of $D$, is within or about a percent level for any short-range
correlation function.

The original iPEPS algorithm needs to assume translational symmetry for
calculation in the thermodynamical limit. The ground state of the
Hamiltonian (1) can spontaneously break the translational symmetry. To take
into account the spontaneous symmetry breaking, we take a large unit cell
and assume the translational symmetry only among different cells with no
symmetry restriction for the tensors within the cell. In our simulation, the
unit cell has $4\times 4$ sites which is large enough to incorporate the
relevant ground states for this Hamiltonian that break the translational
symmetry \cite{note}. We apply imaginary time evolution to reach the ground
state of the Hamiltonian. To avoid being stuck in a metastable state, we
take a number of random initial states for the imaginary time evolution and
pick up the ground state as the one which has the minimum energy over all
the trials.

In Fig. 3, we show the complete phase diagram for the Hamiltonian (1) from
this calculation. To characterize the phase transition, we calculate
derivative of the ground state energy $\frac{\partial E}{\partial J_{2}}%
=\sum_{\left\langle \langle i,j\right\rangle \rangle }\left\langle \vec{S_{i}%
}\cdot \vec{S_{j}}\right\rangle $ with respect to $J_{2}$ ($J_{1}$ is taken
as the energy unit) and identify the singular point of this derivative as
the phase transition point. To characterize properties of different phases,
we calculate the spin order parameter $\left\langle \vec{S_{i}}%
\right\rangle $ for all sites $i$ and the plaquette order parameter $%
Q_{\alpha \beta \gamma \delta }$\cite{14}, defined by

\begin{widetext}
\begin{align}
Q_{\alpha \beta \gamma \delta }& =2\left[ \left\langle \vec{S_{\alpha }}%
\cdot \vec{S_{\beta }}\right\rangle \left\langle \vec{S_{\gamma }}\cdot \vec{%
S_{\delta }}\right\rangle +\left\langle \vec{S_{\alpha }}\cdot \vec{%
S_{\delta }}\right\rangle \left\langle \vec{S_{\beta }}\cdot \vec{S_{\gamma }%
}\right\rangle -\left\langle \vec{S_{\alpha }}\cdot \vec{S_{\gamma }}%
\right\rangle \left\langle \vec{S_{\beta }}\cdot \vec{S_{\delta }}%
\right\rangle \right]   \notag \\
& +1/2\left[ \left\langle \vec{S_{\alpha }}\cdot \vec{S_{\beta }}%
\right\rangle +\left\langle \vec{S_{\gamma }}\cdot \vec{S_{\delta }}%
\right\rangle +\left\langle \vec{S_{\alpha }}\cdot \vec{S_{\delta }}%
\right\rangle +\left\langle \vec{S_{\beta }}\cdot \vec{S_{\gamma }}%
\right\rangle \left\langle \vec{S_{\alpha }}\cdot \vec{S_{\gamma }}%
\right\rangle +\left\langle \vec{S_{\beta }}\cdot \vec{S_{\delta }}%
\right\rangle +1/4\right] ,
\end{align}%
\end{widetext}

where $\alpha $, $\beta $, $\gamma $, and $\delta $ denote four spins on a
plaquette as shown in Fig. 1. Different phases are associated with different
characteristic values of these parameters. For instance, a spin ordered
state is characterized by a significant value of $\left\langle \vec{S_{i}}%
\right\rangle $; in contrast, the plaquette valence bond solid state is
characterized by a near-unity $Q_{\alpha \beta \gamma \delta }$ and a
vanishing $\left\langle \vec{S_{i}}\right\rangle $.

In the inserts of Fig. 2, we show the order parameters $\left\langle \vec{%
S_{i}}\right\rangle $ and $Q_{\alpha \beta \gamma \delta }$ as functions of $%
J_{2}/J_{1}$. These order parameters change abruptly at the corresponding
phase transition points, and the points of abrupt change are in agreement
with the singularity points of $\frac{\partial E}{\partial J_{2}}$. The
order parameters and the derivative of the ground state energy both have
finite jumps at the phase transition points, which strongly indicates that
we have two first-order transitions as we increase the ratio $J_{2}/J_{1}$,
first from a Neel ordered state to a plaquette valence bond solid state at $%
J_{2}/J_{1}=0.88$, and then from the valence bond solid state to another
spin-ordered phase (its nature will be discussed below) at $J_{2}/J_{1}=1.11$%
. The possibility of two second order phase transitions with a coexistence
region of the spin and the valence bond solid orders in the intermediate
region has been discussed in the literature \cite{15}. Within the resolution
of our numerical simulation (the resolution is $0.01$ for the ratio $%
J_{2}/J_{1}$ near the phase transition points), we do not find a coexistence
region and the result supports a direct first-order transition.

\begin{figure}[tbp]
\includegraphics[width=9cm]{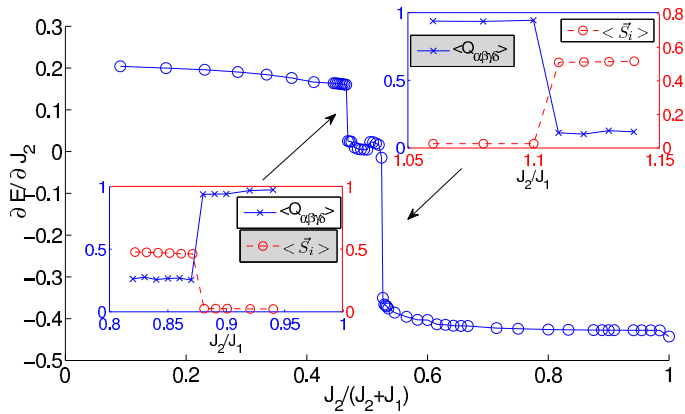}
\caption[Fig. 2 ]{(Color online) The main plot shows $\partial
E/\partial J_{2}$ as a function of $J_{2}/(J_{1}+J_{2})$. 
Insets show the plaquette order (solid line with crosses) and the
spin order (red dashed line with circles) as functions of $J_{2}/J_{1}$%
near the transition points.}
\end{figure}

\begin{figure}[tbp]
\includegraphics[width=8cm]{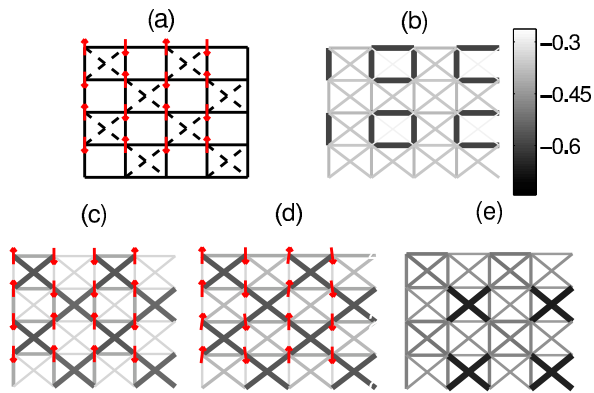}
\caption[Fig. 3 ]{(Color online) The upper left figure (a) shows orientations of
local spins (red arrows) on the checkerboard lattice at $J_{2}=0.5$. The upper
right figure (b) shows nearest neighbor spin-spin correlations $<\vec{S_{i}}\cdot%
\vec{S_{j}}>$ along horizontal, vertical and diagonal bonds on a $16$-site
unit-cell at $J_{2}=1.0$. The width and colors of the bonds are scaled such that
the negative correlation is represented by thicker bonds with darker color. The lower
figures show both spin orientation and correlations $<\vec{S_{i}}\cdot%
\vec{S_{j}}>$ in a Neel{*} state (c)
and a stripe phase (d) at $J_{2}=2$. Fig. (e) illustrates the dimer state, which appears 
as a metastable state at some parameters.}
\end{figure}

The nature of ground states in these three phases are further studied
through calculation of the spin correlation function. In Fig. 3, We show the
nearest-neighbor spin-spin correlation $\left\langle \vec{S_{i}}\cdot \vec{%
S_{j}}\right\rangle $ and orientation of local spins with respect to the
first spin on the up-left corner in each $16$-site unit-cell for three
different $J_{2}$ values. At $J_{2}=1$, strong nearest-neighbor spin
correlations (valence bonds) around the plaquettes breaking the lattice
translational symmetry suggests a plaquette ordered state near this point,
consistent with the finding in Fig. 2. At $J_{2}=0.5$, the spin orientation
indicates a conventional Neel ordered state. At $J_{2}=2$, antiferromagnetic
Neel order appears along the diagonal chains, but not in the horizontal or
vertical axes. With imaginary time evolution starting from randomly chosen
initial states, we actually get two different kinds of spin configurations
shown in Fig. 3(c) and 3(d) for the final state. Their energies are almost
degenerate within resolution of our numerical program. These spin configurations are in agreement with the four-fold
degenerate states found in Ref. \cite{5} based on the large-$S$ expansion
(the other two degenerate states are obtained from Fig. 3(c) and 3(d)
through a $90$-degree rotation of the spin orientation). The spin
configuration in Fig. 3(c) is called the Neel*-state in Ref. \cite{2}, where
the single-site spin $\uparrow $ or $\downarrow $ in the conventional Neel
state is replaced by the two-site unit $\uparrow \uparrow $ or $\downarrow
\downarrow $. The configuration in Fig. 3(d) represents a spin stripe phase,
where the spin orientations form a stripe along the horizontal or vertical
direction, breaking the lattice rotational symmetry. The stripe phase is
also predicted in the large-$N$ calculation \cite{6}.

To further clarify the phase at $J_{2}=2$, we also show the spin-spin
correlation $\left\langle \vec{S_{i}}\cdot \vec{S_{j}}\right\rangle $ in
Fig. 3(c) and 3(d). Strong nearest-neighbor spin correlation appears along
the diagonal chains. However, the distribution of these spin correlations
does not break the symmetry of the lattice, so it is not a cross dimer or
other valence bond solid state. The cross dimer state is predicted for this
model in \cite{2,3} for some region of $J_{2}/J_{1}$. In a cross dimer
state, the nearest-neighbor spin correlations form the cross dimer pattern
illustrated by Fig. 3(e). We indeed get this kind of cross dimer
configuration from the imaginary time evolution starting from a pure cross dimer state
for a certain region of
$J_{2}/J_{1}$ as shown in Fig. 4. However, the energies of the cross dimer
states are strictly higher than the four-fold degenerate states shown in
Fig. 3(c) and 3(d). So the cross dimer state is only a metastable phase in
this region and does not give the real ground state. We compare the energy
of our calculation with the energy of the DMRG calculation in Ref. \cite{3},
and our energy is significantly lower than the corresponding result in \cite%
{3} on the side $J_{2}>J_{1}$. For instance, at $J_{2}=2$, our result shows
a ground state energy of $E=-0.876J_{1}$ for a stripe state, much lower than the energy of $%
E\simeq -0.75J_{1}$ for a cross dimer state at the corresponding point in Ref. \cite{3}. Because of
this large energy difference, it is unlikely that the cross dimer state
emerges as the real ground state of the system.

\begin{figure}[tbp]
\includegraphics[width=8cm,height=3cm]{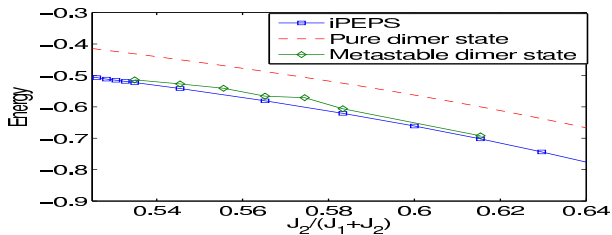}
\caption[Fig. 4 ]{(Color online) Ground state energy calculated by
iPEPS (blue squares) for different $ J_{2}/(J_{2}+J_{1}) $. The dashed line denotes
energy of a pure dimer state, the green diamonds denote energy
of meta-stable dimer states calculated by iPEPS with imaginary time evolution from 
an initial pure dimer state. }
\end{figure}

Some literature predicts a sliding Luttinger liquid state on the $%
J_{2}>J_{1} $ side of the antiferromagnetic checkerboard model \cite{1,1'}.
We do not find evidence to support a transition to a sliding Luttinger
liquid state in our numerical simulation. Of course, due to the limitation
of the internal dimension $D$ of the variational tensor network state in our
calculation, it is possible that the sliding Luttinger liquid state is
poorly approximated by the tensor network state with a small internal
dimension and thus missed in our numerical simulation. We can not rule out
this possibility, however, we feel its chance is small due to the following
test: we know at the limiting point $J_{1}=0$, the model reduces to
decoupled Heisenberg chains whose ground state is described by a Luttinger
liquid with algebraic decay of the spin correlation function. We use the
same tensor network algorithm to calculate the long range spin correlation
for the limiting case at $J_{1}=0$. At this 1D limiting point, we can have a
much larger internal dimension $D$ in numerical simulation, and in Fig. 5(b)
we compare the result with $D$ varying from $2$ to $30$. We see that the
result at $D=5$ has correctly demonstrated the algebraic decay of the spin
correlation function and almost converged to the result at $D=30$. So we do
not necessarily need a large internal dimension for the tensor network
algorithm to uncover the algebraic decay associated with a spin liquid
state. Keeping the same internal dimension at $D=5$, we turn on $J_{1}$ (now
a 2D model with $J_{2}>J_{1}>0$), and find that long range spin
correlation appears along the diagonal chains (see Fig. 5(a)), indicting
that the spin order along this direction is very likely a real effect.

\begin{figure}[tbp]
\includegraphics[width=8cm,height=5cm]{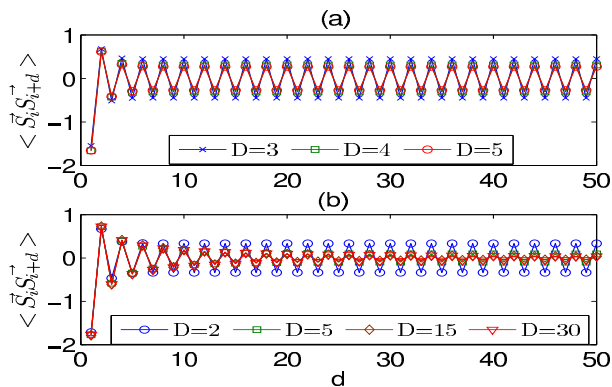}
\caption[Fig. 5 ]{(Color online) (a) Long-range spin
correlations along a diagonal chain obtained with $D=3$ (crosses), $D=4$ 
(squares), and $D=5$ (circles) at $J_{2}=2$. (b) The same
correlation for a Heisenberg chain calculated with $D=3$ (circles), 
$D=5$ (squares), $D=15$ (diamonds), and $D=30$ (triangles). }
\end{figure}

Although our numerical program can not distinguish the energy of the four
degenerate states shown in Fig. 3(c,d) at the $J_{2}>J_{1}$ side, it is very
likely that the stripe phase will emerge as the real ground state in
practice because of its robustness to perturbation in the Hamiltonian. In
real realization of the model Hamiltonian (1), the $J_{1}$ coupling along
the horizontal and the vertical directions might be slightly different, or
apart from the $J_{2}$ coupling on the checkerboard pattern, there might be
small antiferromagnetic $J_{2}$ coupling along the other plaquettes. With
any of these types of perturbation (which sound to be inevitable in
practice), the energy of the Neel*-state will be lifted, and the stripe
phase will emerge as the unique ground state of the system.

In summary we have used the iPEPS method, a type of tensor networks
algorithms, to calculate the ground states of the frustrated
anti-ferromagnetic $J_{1}$-$J_{2}$ Heisenberg model on a checkerboard
lattice. We construct the complete phase diagrams, indicting two first order
phase transitions, first from a Neel state to a plaquette valence bond solid
and then to a spin stripe phase. The calculation helps to clarify some of
the previous debates on the phase diagram of this important model and
provides a novel example for applications of the recently developed tensor
network algorithms to frustrated systems.

We thank Subir Sachdev and Chao Shen for helpful discussions. This work was
supported by the NBRPC (973 Program) 2011CBA00300 (2011CBA00302), the DARPA
OLE program, the IARPA MUSIQC program, the ARO and the AFOSR MURI program.


\begin{thebibliography}{99}
\bibitem{note} A finite unit cell could suppress incommensurate spin orders,
although such incommensurate orders do not sound to be a likely candidate
for the ground state of the Hamiltonian (1).
\end{thebibliography}
\end{document}